\documentclass[12pt]{article}
\usepackage{epsf}
\usepackage{amsmath}
\usepackage{graphics}
\usepackage{cite}

\renewcommand{\thefootnote}{\#\arabic{footnote}}
\begin{document}

\newcommand{\gtrsim}{ \mathop{}_{\textstyle \sim}^{\textstyle >} }
\newcommand{\lesssim}{ \mathop{}_{\textstyle \sim}^{\textstyle <} }

\newcommand{\rem}[1]{{\bf #1}}

\renewcommand{\thefootnote}{\fnsymbol{footnote}}
\setcounter{footnote}{0}
\begin{titlepage}

\def\thefootnote{\fnsymbol{footnote}}

\begin{center}
\hfill astro-ph/0608138\\
\hfill August 2006\\
\hfill Preliminary Version\\
\vskip .5in
\bigskip
\bigskip
{\Large \bf Deflation at Turnaround for Oscillatory Cosmology}

\vskip .45in

{\bf Lauris Baum and Paul H. Frampton}

{\em Institute of Field Physics, Department of Physics and Astronomy,}

{\em University of North Carolina, Chapel Hill, NC 27599-3255, USA}

\end{center}

\vskip .4in
\begin{abstract}
It is suggested that dark energy in a brane world can help 
reconcile an infinitely cyclic cosmology with the second 
law of thermodynamics. A cyclic cosmology is described, 
in which dark energy with constant equation of state
leads to a turnaround at finite future time, when entropy
is decreased by a huge factor equal to the inverse
of its enhancement during the initial inflation. 
Thermodynamic consistency of cyclicity requires
the arrow of time to reverse during contraction.
Entropy reduction in the contracting phase
is infinitesimally smaller than entropy increase
during expansion.

\end{abstract}
\end{titlepage}

\renewcommand{\thepage}{\arabic{page}}
\setcounter{page}{1}
\renewcommand{\thefootnote}{\#\arabic{footnote}}

\newpage

\bigskip

\noindent {\it Introduction}

\bigskip

One of the oldest questions in theoretical cosmology is whether
an infinitely oscillatory universe which avoids an initial singularity
can be consistently constructed. As realized by Friedmann\cite{Friedmann}
and especially by Tolman\cite{Tolman,TolmanBook} one principal
obstacle is the second law of thrmodynamics which dictates that
the entropy increases from cycle to cycle. If the cycles 
thereby become longer, extrapolation into the past
will lead back to an initial singularity again, thus removing the
motivation to consider an oscillatory universe in the first place.
This led to the abandonment of the oscillatory universe by
the majority of workers.

Nevertheless, an oscillatory universe is an attractive alternative
to the Big Bang. One new ingredient in the cosmic make-up
is the dark energy discovered only in 1998 and so
it natural to ask whether this new component in
the Friedmann equation can avoid the difficulties with entropy
which have dogged previous attempts.

Some work has been started to exploit the dark energy
in allowing cyclicity possibly without apparently
the need for inflation in
\cite{SteinhardtTurok,SteinhardtTurok2,Boyle,SteinhardtTurok3}.
Another new ingredient is the use of branes and a fourth spatial
dimension as in \cite{Sundrum,Randall,Binetruy,Freese} which
have examined the consequences for cosmology. The Big Rip
and replacement of dark energy by modified gravity have
been explored in \cite{PHFTT,PHFTT2}.

If the dark energy has a super-negative equation of state,
$\omega_{\Lambda} = p_{\Lambda}/\rho_{\Lambda} < -1$, it leads
to a Big Rip at a finite time where there
exist extraordinary conditions with regard to
density and causality as one approaches the Big Rip.
In the present article we explore whether these exceptional
physical conditions can assist in providing a truly
infinitely-cyclic entropy density
in an oscillatory universe of time periodicity $t=t$ ( mod $\tau$).

We shall consider the situation where if, as we approach the Big Rip, 
the expansion stops just short of the rip and there is a turnaround
at $t=t_T$ (mod $\tau$ ) when the scale factor is deflated to 
a very tiny fraction ($f$ in our notation) 
of itself. For the deflation there is a consistency condition,
written in Eq.(\ref{deflation}).
Entropy is extensive so a fraction $(1-f^3)$ is jettisoned
at turnaround. One key ingredient is that the turnaround takes
place a sufficiently short time before the Big Rip would
have occurred, at a time when the universe is fractionated into many
causal patches \cite{PHFTT2}.

We then proceed to investigate the contraction phase which
occurs with a very much smaller universe than in
the expansion phase. 
A bounce at $t=\tau$ ( mod $\tau$) takes place 
a short time before what
would have been the Big Bang. Then, immediately after the bounce, entropy
is injected as usual by inflation\cite{Guth} where the scale factor
is enhanced by factor $E$ and hence entropy by $E^3$. Inflation is
thus an essential part of the present scenario which is
one distinction from the work of 
\cite{SteinhardtTurok,SteinhardtTurok2,Boyle,SteinhardtTurok3}.  

For cyclicity of the entropy, $S(t) = S(t + \tau)$ to be consistent with thermodynamics 
it is insufficient that the huge inflationary enhancement $E^3$ be
completely compensated by deflation at turnaround.
Additionally, it is necessary for the thermodynamic arrow of time to reverse
during contraction. This is one shortcoming of the proposal.
The decrease in entropy during contraction is 
infinitesimal, being at most $10^{-84}$ (sic) of the entropy increase during
expansion.
The parameters $f$ and $E$ are related by
consistency of the expansion and contraction.

A second possible shortcoming of the proposal is the persistence of 
spacetime singularities in cyclic cosmologies\cite{Vilenkin} which
we do not address. 

This paper is published because
our discussion seems to give 
a plausible realization of the oscillatory universe originally
saught in \cite{Friedmann,Tolman,TolmanBook} whose shortcomings can
hopefully be evolved by others into a more convincing scenario.
The shortcomings are discussed 
again in the Final Discussion section.

\bigskip
\bigskip
\bigskip

\noindent {\it Friedmann Equation for Expansion phase}

\bigskip
\bigskip

Let the period of the Universe be designated by $\tau$
and the bounce take place at $t = 0$ ( mod $\tau$)
and turnaround at $t = t_T$ ( mod $\tau$).
Thus the expansion phase is for times $0 < t < t_T$ ( mod $\tau$)
and the contraction phase corresponds to times
$t_T < t < \tau$ ( mod $\tau$).

\bigskip

We shall employ the following Friedmann equation for
the {\it expansion} period $0 < t < t_T$ ( mod $\tau$):

\begin{equation}
\left( \frac{\dot{a}(t)}{a(t)} \right)^2  =   
\frac{8 \pi G}{3} \left[ \left( \frac{(\rho_{\Lambda})_0}{a(t)^{3(\omega_{\Lambda} + 1)}}
+\frac{(\rho_{m})_0}{a(t)^{3}} +\frac{(\rho_{r})_0}{a(t)^{4}}
\right)
-\frac{\rho_{total}(t)^2}{\rho_c}
\right] 
\label{Friedmann}
\end{equation}

\bigskip

\noindent or, more succinctly,

\bigskip
\begin{equation}
\left( \frac{\dot{a}(t)}{a(t)} \right)^2  =   
\frac{8 \pi G}{3} \rho_{total} (t) \left[
1  -\frac{\rho_{total}(t)}{\rho_c} \right]
\end{equation}
where the scale factor is normalized to $a(t_0)=1$ at the present time
$t = t_0 \simeq 14Gy$. 

\bigskip

To explain the notation, $(\rho_i)_0$ denotes the value of the density $\rho_i$
at time $t=t_0$. The first two terms are the dark energy and total matter
(dark plus luminous) satisfying
\begin{equation}
\Omega_{\Lambda} = \frac{8 \pi G (\rho_{\Lambda})_0}{3 H_0^2} = 0.72 
~~~ {\rm and} ~~~
\Omega_{m} = \frac{8 \pi G (\rho_m)_0}{3 H_0^2} = 0.28 
\end{equation}
where $H_0 = \dot{a}(t_0)/a(t_0)$. The third term in the Friedmann equation
is the radiation density which is now 
$\Omega_r = 1.3 \times 10^{-4}$.
The final
term $\sim \rho_{total}(t)^2$ is derivable from the
Randall-Sundrum set-up\cite{Sundrum,Randall,Freese}; we use
a negative sign arising either from a timelike extra dimension
or, preferably, from a negative brane tension.
$\rho_{total} = \Sigma _{i=\Lambda, m, r} \rho_{i}$.
As the turnaround is approached, the only significant terms 
in Eq.(\ref{Friedmann}) are the
first (where $\omega_{\Lambda} < -1$) and the last. As the bounce is approached,
the only important terms in Eq.(\ref{Friedmann}) are the third and the last.

\bigskip

In particular, the final term of Eq. (\ref{Friedmann}), $\sim \rho_{total}(t)^2$, 
arising from the brane set-up is insignificant
for almost the entire cycle but becomes dominant
as one approaches $t \rightarrow t_T$ ( mod $\tau$ ) for the turnaround
and again for $t\rightarrow \tau$ ( mod $\tau$ ) approaching the bounce.

\bigskip
\bigskip

\noindent {\it Turnaround Compared to  Big Rip}

\bigskip
\bigskip

Let us assume for algebraic simplicity
$\omega_{\Lambda} = -4/3 = {\rm constant}$. This
value is already almost excluded by WMAP3 \cite{WMAP3} but
to begin we are not aiming not at realistic cosmology but at
consistency of infinite cyclicity. More realistic values
will be discussed elsewhere.

\bigskip

The approach to the Big Rip will follow that discussed in
\cite{PHFTT,PHFTT2}. With the value $\omega_{\Lambda}=-4/3$
we learn from \cite{PHFTT} that the time to the Big Rip
is $(t_{rip}-t_0) = 11 {\rm Gy} (-\omega_{\Lambda} - 1)^{-1} = 33 {\rm Gy}$
which is, within one second,  when turnaround occurs at $t=t_T$.
So if we adopt $t_0=14 Gy$ then $t_T = t_0 + (t_{rip}-t_0) = (14 + 33) Gy = 47 Gy$. 

\bigskip

From the analysis in \cite{PHFTT,PHFTT2} 
the time when a system becomes graviatationally unbound corresponds approximately
to the time when the growing dark energy density matches the mean density
of the bound system. For a ``typical" object like the Earth (or a hydrogen atom where
the mean density happens to be about the density of water $\rho_{H_2O}=1g/cm^3$ 
since $10^{-24}g/(10^{-8}cm)^3 = 1g/cm^3$) water's density $\rho_{H_2O}$ 
is an unlikely but practical unit for cosmic density in the oscillatory
universe. 

\bigskip

With this in mind, for the simple case of $\omega=-4/3$ we see from
Eq.(\ref{Friedmann}) that the dark energy density grows proportional
to the scale factor $\rho_{\Lambda}(t) \propto a(t)$ and so given
that the dark energy at present is $\rho_{\Lambda} \sim 10^{-29} g/cm^3$
it follows that $\rho(t_{H_2O}) = \rho_{H_2O}$ when $a(t_{H_2O}) \sim 10^{29}$. 

\bigskip

So the next step, equally straightforward, is to estimate the time $t_{H_2O}$
when this occurs.
If we take the Friedmann equation with only dark energy
\begin{equation}
\left( \frac{\dot{a}}{a} \right)^2 = H_0^2 \Omega_{\Lambda} a^{-\beta}
\label{DEequation}
\end{equation}
with $\beta=3(1+\omega)$, then the scale factor evolves as
\begin{equation}
a(t)= \left[ \frac{2}{3 H_0 \sqrt{\Omega_{\Lambda}} 
(-1 - \omega) (t_{rip} - t)}\right]^{-\frac{2}{3(1+\omega)}}
\label{scale}
\end{equation}
and, for later use, it follows that 
\begin{equation}
\frac{a(t_1)}{a(t_2)} = \left( \frac{(t_{rip}-t_2)}{(t_{rip}-t_1)}
\right)^{-\frac{2}{3(1+\omega)}}
\label{a1a2}
\end{equation}

\bigskip

When we again specialize to $\omega= - 4/3$ as illustration
and require as before $\rho_{\Lambda}(t_{H_2O}) = \rho_{H_2O}$
then $a(t_{H_2O})= 10^{29}$ and from Eq.(\ref{a1a2})
it follows that 
\begin{equation}
\frac{a(t_{H_2O})}{(a(t_0)=1)} = \left( \frac{(t_{rip}-t_0)}
{(t_{rip}-t_{H_2O})}
\right)^{2}
\label{aUa0}
\end{equation}
so that $(t_{rip}-t_{H_2O}) = 33Gy \times 10^{-14.5}
\simeq 10^{3.5}s \sim 1$ hour.

\bigskip

\noindent [These values are sensitive to $\omega$: if we choose
$\omega=-1.1$ the corresponding results
are $(t_{rip}-t_0) = 110Gy$ and $(t_{rip}-t_U) = 5.25 My$.] 

\bigskip

\noindent Returning to the case $\omega=-4/3$, it will be useful to consider
a more general critical density $\rho_c = \eta \rho_{H_2O}$, since
there is nothing special about $\rho_{H_2O}$ and to
compute the time $(t_{rip} - t_{\eta})$ such that
$\rho_{\Lambda} (t_{\eta}) = \rho_c =  \eta \rho_{H_2O}$.
Similarly to Eq. (\ref{aUa0}), we then find
\begin{equation}
\frac{a(t_{\eta})}{(a(t_0)=1)} = \left( \frac{(t_{rip}-t_0)}
{(t_{rip}-t_{\eta})}
\right)^{2}
\label{aetaa0}
\end{equation}
That is, using $a(t_{\eta}) = 10^{29} \eta$,
\begin{equation}
(t_{rip} - t_{\eta}) = (t_{rip}-t_0) 10^{-14.5} \eta^{-1} \simeq \eta^{-1} {\rm hours}
\label{teta}
\end{equation}

\bigskip

\noindent which is the required result.

\bigskip
\bigskip

\noindent {\it Deflation at Turnaround.}

\bigskip
\bigskip

A key ingredient in our cyclic model is that at turnaround
$t = t_T$ ( mod $\tau$ ) our universe deflates dramatically
with scale factor $a(t_T)$ shrinking to $\hat{a}(t_T) = f a(t_T)$
where $f < 10^{-28}$. 

This jettisoning of almost all, a fraction $(1-f)$, of the
accumulated entropy is permitted by the exceptional causal
structure of the universe. We shall show later that in Eq.
(\ref{teta}) the parameter $\eta$ satisfies $\eta > 10^{27}$
(see Eq.(\ref{eta})) which implies the unimaginable dark energy density at turnaround
of $\rho_{\Lambda}(t_T)  > 10^{27}\rho_{H_2O}$. By the time the dark energy
density reaches such a value, according
to the Big Rip analysis of \cite{PHFTT,PHFTT2} the smallest
known bound systems of particles have become unbound.
Additionally the constitutents will be causally disconnected
meaning that if the expansion had, instead, continued to the Big Rip the
particles could no longer causally communicate \cite{PHFTT2}.

The Hubble parameter $H$ for deflation at $t = t_T$ ( mod $\tau$ ) satisfies
\begin{equation}
\int_{a(t_T)}^{\hat{a}(t_T)} H da = - \int_{\hat{a}(\tau)}^{a(\tau)} H da 
\label{deflation}
\end{equation}
where the right hand side refers to inflation at $t = \tau$ ( mod $\tau$ ).
We do not discuss here the brane dynamics leading to Eq.(\ref{deflation})
but such a condition 
\footnote{Presumably the complete five-dimensional brane-world
calculation will reveal that as the Big Rip is approached a transition 
$a(t_T) \rightarrow \hat{a} (t_T)$ is dictated by energy considerations. 
We do not pursue the calculation here but merely suggest the result 
of a more complete treatment.}
is necessary for true periodicity.

\bigskip
\bigskip

\noindent {\it Friedmann Equation for Contraction Phase.}

\bigskip
\bigskip

\noindent The contraction phase occurs for the period $t_T < t < \tau$ ( mod $\tau$ ).
The scale factor for the contraction phase will be denoted by $\hat{a} (t)$
while for time we use always the same linear time $t$ subject to the
periodicity $t + \tau \equiv t$.

\bigskip

At the turnaround we retain a fraction $f^3$ of the entropy with
$\hat{a}(t_T) = f a(t_T)$ and for the contraction phase the Friedmann equation
is
\begin{equation}
\left( \frac{\dot{\hat{a}}(t)}{\hat{a}(t)} \right)^2  =   
\frac{8 \pi G}{3} \left[ \left( \frac{(\hat{\rho}_{\Lambda})_0}{\hat{a}(t)^{3(\omega_{\Lambda} + 1)}}
+\frac{(\hat{\rho}_{m})_0}{\hat{a}(t)^{3}} +\frac{(\hat{\rho}_{r})_0}{\hat{a}(t)^{4}} \right)
-\frac{\hat{\rho}_{total}(t)^2}{\rho_c}
\right] 
\label{hatFriedmann}
\end{equation}

\noindent where we have defined
\begin{equation}
\hat{\rho}_i(t) = 
\frac{(\rho_i)_0 f^{3(\omega_i +1)}}{\hat{a}(t)^{3(\omega_i+1)}}  
= \frac{(\hat{\rho}_i)_0}{\hat{a}(t)^{3(\omega_i+1)}}  
\label{hatrho}
\end{equation}

\bigskip

\noindent Let us define landmarks in the contraction phase at times $t^{'}_0$ and $t^{'}_{eq}$ defined by
obvious analogy with expansion

\begin{equation}
\hat{\rho}_{\Lambda}(t^{'}_0) = \hat{\rho}_m(t^{'}_0) (\rho_{\Lambda})_0/(\rho_{m})_0
\label{tprime0}
\end{equation}
and
\begin{equation}
\hat{\rho}_{r}(t^{'}_{eq}) = \hat{\rho}_m(t^{'}_{eq})
\label{tprimeeq}
\end{equation}

\noindent The period $\tau$ will be set by $\rho_c$ = maximum 
($\hat{\rho}_i(\tau)$)$_{i=\Lambda, m, r}$. The maximum will
generally be for the most positive $\omega_i$.

\bigskip

\noindent For the contraction from $t=t_T$ to the time $t=t_0^{'}$ the Friedmann equation is dominated
by the first term on the R.H.S. of Eq.(\ref{hatFriedmann}):

\begin{equation}
\dot{\hat{a}}(t) = - \sqrt{\frac{8 \pi G (\hat{\rho}_{\Lambda})_0}{3}} ~~ \hat{a}^{3/2}
\label{DEcontraction}
\end{equation}

\bigskip

\noindent Integration of Eq. (\ref{DEcontraction}) from $t=t_T$ to $t=t_0^{'}$ 
and using $\hat{a}(t_0^{'}) \ll \hat{a}(t_T)$ gives

\begin{equation}
\hat{a}(t_0^{'}) = \frac{4 f}{\Omega_{\Lambda} H_0^2} \frac{1}{(t_0^{'} - t_T)^2}
=f
\label{hata0}
\end{equation}

\bigskip

\noindent Using the definition of $t_0^{'}$ from Eq. (\ref{tprime0}) results in
$\hat{a}(t_0^{'})^4 = f^4 (\rho_m)_0/(\rho_{\Lambda})_0$ and 

\begin{equation}
(t_0^{'} - t_T) \simeq \frac{2}{(\sqrt{\Omega_{\Lambda}}) H_0} \sim 33 Gy 
\label{tTt0}
\end{equation}
using $H_0 = 70 km/s/Mpc$ and $1Mpc=3.1\times10^{19}km$, hence $H_0^{-1}\simeq 14Gy$.

\bigskip
\bigskip

\noindent The second stage of the contraction is from $t = t_0^{'}$ (mod $\tau$ ) to
$t = t_{eq}^{'}$ ( mod $\tau$ ) controlled by

\begin{equation}
\dot{\hat{a}}(t) = - \sqrt{\frac{8 \pi G (\hat{\rho}_{m})_0}{3}} ~~ \hat{a}^{-1/2}
\label{DMcontraction}
\end{equation}

\bigskip

\noindent Integration of Eq. (\ref{DMcontraction}) from $t=t_0^{'}$ to $t=t_{eq}^{'}$ 
and using $\hat{a}(t_{eq}^{'}) \ll \hat{a}(t_0^{'})$ gives

\begin{equation}
\hat{a}(t_{eq}^{'}) = \frac{(\hat{\rho}_r)_0}{(\hat{\rho}_m)_0} = 5 \times 10^{-4} f 
\label{hataeq}
\end{equation}

\bigskip

\noindent Using the definition of $t_{eq}^{'}$ from Eq. (\ref{tprimeeq}) results in

\begin{equation}
(t_{eq}^{'} - t_{0}^{'}) \simeq \frac{2}{3 (\sqrt{\Omega_{m}}) H_0} \sim 13 Gy 
\label{t0teq}
\end{equation}

\bigskip
\bigskip

The third and final stage of contraction is from
$t=t_{eq}^{'}$ ( mod $\tau$ ) to the bounce at time $t = \tau$ 
( mod $\tau$ ) wich is radiation dominated with

\begin{equation}
\dot{\hat{a}}(t) = - \sqrt{\frac{8 \pi G (\hat{\rho}_{r})_0}{3}} ~~ \hat{a}^{-1}
\label{rcontraction}
\end{equation}

\noindent Integration of Eq. (\ref{rcontraction}) from $t=t_{eq}^{'}$ to $t= \tau$ 
and using $\hat{a}(\tau) \ll \hat{a}(t_{eq}^{'})$ gives

\begin{equation}
(\tau - t_{eq}^{'}) = \frac{1.25 \times 10^{-7}}
{(\sqrt{\Omega_r}) H_0} \simeq 150,000 y
\label{ttauteq}
\end{equation}

\bigskip
\bigskip

\noindent {\it Bounce at $t=\tau$ ( mod $\tau$ )}

\bigskip

As a estimated time $t = \tau$ ( mod $\tau$ ) for the bounce, 
the contraction scale is given, using $\rho_c = \eta \rho_{H_2O}$,
from Eq. (\ref{Friedmann}) as

\begin{equation}
a (\tau)^4 = \left( \frac{ (\rho_r)_0}{\eta \rho_{H_2O}}
\right) \simeq \left( \frac{10^{-33}}{\eta} \right)
\label{atau}
\end{equation}

Now the bounce at $t=\tau$ ( mod $\tau$ ) must be after
the Planck time $t_{Pl}=10^{-44}s$ when $a( t_{Pl}) \sim 10^{-32}$
nd before the electroweak transition at
$t_{EW}=10^{-10}s$ when $a(t_{EW}) = 10^{-15}$. From
Eq.(\ref{atau}) this imples for $\eta$ that

\begin{equation}
10^{27} < \eta < 10^{95}
\label{eta}
\end{equation}

\bigskip
\bigskip

The parameter $f=\hat{a}(t_T)/a(t_T)$ defined at turnaround
reappears at the bounce by $\hat{a}(\tau) = f a(\tau)$.
Immediately after the bounce there is conventional
inflation with enhancement E = $a(\tau)/\hat{a}(\tau)$
and successful inflation requires $E > 10^{28}$. Consistency
requires Ef=1 and therefore $f < 10^{-28}$. The fraction
of entropy jettisoned in deflation immediately
after turnaround  is at least $(1 - 10^{-28})$.

\bigskip
\bigskip

\noindent {\it Four Discussions of Entropy}

\bigskip
\bigskip

\noindent (i) Entropy at present $t=t_0$ ( mod $\tau$ ).

\bigskip

\noindent Having set up an oscillatory universe by massaging 
the four terms in our Friedmann equation, Eq.(\ref{Friedmann}),
it behooves us to discuss the entropy at different epochs
in the cycle,  with a view to find the minimal set of
assumptions necessary to reconcile cyclicity with the
second law of thermodynamics. 

\bigskip

Consider first the present epoch $t=t_0$. The contributions
of the radiation to the entropy density $s$ follows
the relation
\begin{equation}
s = \frac{2 \pi^2}{45}g_* T^3
\label{entropy}
\end{equation}
First consider only photons with $g_*=2$. The present
CMB temperature is $T=2.73K \equiv 0.235 meV \sim 1.191 (mm)^{-1}$. 
Substitution in Eq.(\ref{entropy}) gives a present radiation 
entropy density $s_{\gamma}(t_0)= 1.48 (mm)^{-3}$. Using a volume estimate
$V=(4\pi/3)R^3$ with $R=0Gly \simeq \times10^{29}mm$ gives a total
radiation entropy
$S_{\gamma} \sim 6.3 \times 10^{87}$. Including neutrinos increase $g_*$ in
Eq.(\ref{entropy}) from $g_*=2$ to $g_*=3.36=2+6\times(7/8) \times (4/11)^{4/3}$.
This increases $S_{\gamma}=6.3 \times 10^{87}$ to $S_{\gamma+\nu} \sim \times 10^{88}$. 
                                   
This total entropy is interpretable as $exp (10^{88})$ degrees of freedom,
or in information theory\cite{KN} to a number $I$ of qubits where
$2^I = e^S$ so that $I = S/(ln 2 = 0.693) \sim 10^{88}$. This is well below
the holographic bound which is dictated by the area in terms
of Planck units $10^{-64} mm^2$ which gives $S_{holog}(t_0) = 4\pi (10^{29} mm)^2
/(10^{-32} mm)^2 \sim 10^{123}$ about $10^{35}$ times bigger.
In \cite{KN} it is suggested that some of this difference may
come from supermassive black holes.

The entropy contribution from the baryons is smaller than $S_{\gamma}$
by some ten orders of magnitude, so like that of the dark matter,
is negligible. 

What is the entropy of the dark energy? If it is perfectly homogeneous
and non-interacting it has zero temperature and entropy: this is 
our assumption here. Another viewpoint, at least for a pure cosmological
constant, is that one number $\Lambda$ cannot contain entropy.

Finally, the 4th term in Eq.(\ref{Friedmann}) corresponding
to the brane term is neglible, as we have already estimated. 

\bigskip

The conclusion is that $S_{total}(t_0) \sim 10^{88}$.

\bigskip
\bigskip

\noindent (ii) Entropy at turnaround $t=t_T$ ( mod $\tau$ ).

\bigskip
\bigskip

We have estimated that $a(t_T) = 10^{29} \eta$. The temperature
$T_{\gamma}$ of the radiation scales as $T_{\gamma} \propto a(t)^{-1}$
so using the entropy
density of Eq.(\ref{entropy}) a comoving 3-volume 
$\propto a(t)^3$ will contain
the same total radiation entropy $S_{\gamma}(t_T) = S_{\gamma}(t_0)$
as at present; this is simply the usual adiabatic expansion.

\bigskip

The expansion from $t=0$ ( mod $\tau$) to $t_T$ ( mod $\tau$ ) 
is, of course, not purely adiabatic
because irreversible processes take place. There are
phase transitions such as  
the electroweak transition at $t_{ew}\sim 100ps$, the
QCD phase transition at $t_{QCD} \sim 100 \mu s$, 
and recombination at $t_{rec}\sim 10^{13}s$. Further
irreversible processes occur during during stellar evolution.
Although the expansion of the radiation, the
dominant contributor to the entropy is
close to adiabatic, the entropy of the matter
inevitably increases with time in accord with the
second law of thermodynamics.

\bigskip

In our model, the entropy of the matter increases
between $t = 0$ ( mod $\tau$ ) and $t_T=47Gy$ ( mod $94 Gy$ ). 
Even approximating the entropy of the dark energy as zero and the
radiation as adiabatic, the matter part represented
by $\rho_m$ will cause the entropy to rise
from $S(t = 0)$ to $S(t_T) = S(t = 0) + \Delta S$ where $\Delta S$
causes the contradiction plaguing the oscillatory
universe for a long time \cite{Friedmann,Tolman,TolmanBook}.

\bigskip

The key point is that in order for entropy 
to be cyclic, the entropy which was enhanced by a huge
factor $E^3 > 10^{84}$ at inflation must be 
reduced equally dramatically at some point during the cycle
so that $S(t) = S(t+\tau)$ becomes possible. Our proposal 
is that entropy is so reduced, or deflated, 
at the turnaround $t=t_T$ by jettisoning causally disconnected
regions\cite{PHFTT2} and keeping only a fraction
$f^3 < 10^{-84}$ of the entropy. The second law of thermodynamics
continues to obtain but other causally disconnected regions are 
permanently removed from our universe at each turnaround.

\bigskip

\noindent (iii) Entropy for contraction $t_T < t < \tau$ ( mod $\tau$ )

\bigskip

The discussion of the entropy is the most interesting during 
the contraction phase. According to statistical mechanics one
expects the entropy to change from 
$\hat{S}(t_T) = 10^{-84}[S(t=0) + \Delta S]$ to
$\hat{S}(\tau) = \hat{S}(t_T) + \Delta \hat{S}$ and that
$\Delta \hat{S}$ be positive. At least that is inevitable
if the thermodynamic arrow of time remains in the direction
of positive $t$. 

\bigskip

There are immediate difficulties. Consider just one of the
transitions, the recombination in reverse. For the contracting
universe confronting such a phase transition will prematurely
bounce if $\Delta \hat{S} > 0$ s maintained. 

\bigskip

A consistent possibility that will permit the
tiny universe to execute the required transitions
in reverse is that 

\begin{equation}
\Delta \hat{S} = - 10^{-84} \Delta S 
\label{condition}
\end{equation}
and then premature bouncing can be avoided.

\bigskip

A way to implement this is to reverse the thermodynamic
arrow of time during contraction. We consider this one of the two
principal weaknesses of the present picture (see the Final
Discussion) but such an
assumption is necessary to allow contraction all the way
from $t=t_T$ ( mod $\tau$ ) to $t=\tau$ ( mod $\tau$ ). 

\bigskip

Because the contracting universe is so small, only
a decrease by $\Delta \hat{S}$ given by
Eq. (\ref{condition}) is required, but this
question requires clarification. We have nothing
noteworthy to add.

\bigskip

\noindent (iv) Entropy at bounce $t=\tau$ (mod $\tau$ )

\bigskip

Immediately after turnaround the inflation increases
entropy by $10^{84}$ so $S(t)=S(t+\tau)$ providing
Eq.(\ref{condition}) or its alternative is satisfied.
The overall symmetry between the initial inflation
and the deflation at turnaround are one appealing
aspect of this particular version of a cyclic
universe.

\bigskip
\bigskip

{\it Final Discussion}

\bigskip
\bigskip

\noindent The standard cosmology based on a Big Bang augmented by
an inflationary era is impressively consistent with the detailed
data from WMAP3 \cite{WMAP3} when dark energy, most conservatively
a cosmological constant, is included.

\bigskip

\noindent The objections to this standard model
are largely philosophical and not motivated directly by observations.
The first objection is the nature of the initial singularity and the initial
conditions. A second objection, not of concern to some colleagues, is that 
the predicted fate of the universe is an infinitely long expansion.

\bigskip

We have attempted to outline a cosmology where these objections
are answered. But the proposal has, itself, shortcomings. The two
most serious are in our opinion:

\begin{itemize}

\item The reversal of the thermodynamic arrow of time during the contraction
phase renders our proposal dubious. The fact that the necessary entropy
reduction is at most $10^{-84}$ of the entropy gain during expansion
does not address the deeper issue of how entropy decreases.

\item One motivation for a cyclic cosmology is the avoidance of
spacetime singularities. With regard to
the powerful results of \cite{Vilenkin}, their theorem
is not immediately applicable because it assumes 
the average expansion parameter 
\footnote{Given that a quasi-static universe
is untenable for reasons of instability, a corollary of the theorem
proved in \cite{Vilenkin} is that the {\it only} possible
infinite lifetime universe is one where $H_{av}=0$.}
is $H_{av} > 0$ whereas here $H_{av}=0$. However,
this may provide only temporary relief from an ubiquitous difficulty.

\end{itemize}

\bigskip

We publish the present deflationary proposal mainly in the hope that
it will stimulate an improved and more consistent formulation by
others.

\bigskip
\bigskip
\bigskip
\bigskip
\bigskip

\begin{center}

{\bf Acknowledgements}

\end{center}

\bigskip
We thank Alex Vilenkin for useful discussion about paper \cite{Vilenkin}.
This work was supported in part by the
U.S. Department of Energy under Grant No. DE-FG02-97ER-41036.

\end{document}